\documentclass[
    ,final            
  ,mathptm
]{aipproc}

\layoutstyle{6x9}

\begin{document}

\title{DIM light on Black Hole X-ray Transients}

\classification{97.80.Jp, 98.62.Mw}
\keywords      {X-ray binaries; Infall, accretion, and accretion disks}

\author{Guillaume Dubus}{
address={Laboratoire Leprince-Ringuet, UMR 7638 CNRS, Ecole Polytechnique, 91128 Palaiseau France}
,address={Institut d'Astrophysique de Paris, UMR 7095 CNRS, UPMC, 98 bis bd Arago, 75014 Paris France}
}

\begin{abstract}
The current model for the outburst cycles of stellar-mass black holes X-ray binaries is the {\em disk instability model} (DIM). An overview of this model and a discussion of its theoretical and observational challenges are given.
\end{abstract}

\maketitle


Most known low-mass X-ray binaries (LMXBs) are transient, their radio to X-ray emission changing on timescales of days-months by orders of magnitude \cite{ccl}. The X-ray brightening can be spectacular as some transients briefly outshine the brightest X-ray sources before returning to low quiescence levels. 

Unlike neutron star LMXBs, all known black holes with low-mass companions are transients. Thin accretion disks are unstable when the companion mass transfer rate is low. Why BH LMXBs have low $\dot{M}_t$ may have to do with binary evolution,  but this is hard to ascertain due to the scarcity of LMXBs with which scenarios can be tested (an already difficult project with the more abundant cataclysmic variables). \S1 recaps the caveats, successes and shortcomings of the {\em disk instability model} (DIM) proposed to explain the outbursts. \S2 examines some of the theoretical and observational challenges.

\section{The Disk Instability Model}
The outburst mechanism in transient LMXBs is thought to be the same as in cataclysmic variables \cite[for a review see][]{lasotarev}. The accretion flow is a thin disk with $\approx$ Keplerian velocities and $c_s/R\Omega_K\sim H/R\ll 1$. Angular momentum transport is likely mediated by the magneto-rotational instability (MRI). For a thin, weakly turbulent disk, the effect can be captured by a stress tensor with a dominant $R\phi$ component proportional to pressure, providing some justification for the $\alpha$ parametrization of transport used in the DIM \cite{balbuspap}.

Energy released by angular momentum transport (`viscous dissipation') at each radius in a thin disk is locally balanced by radiative losses. Only a certain set of temperatures and densities satisfy the thermal balance requirement for given $R$ and $\alpha$. The opacity changes rapidly when hydrogen ionises, influencing the thermal equilibrium. This regime ($T\sim 7000~$K) turns out to be thermally (and viscously) unstable. The result depends little on the assumptions made in the calculation (vertical energy dissipation, radiative transfer or magnitude of $\alpha$).

A stationary, thin disk has a temperature profile $T^4\propto \dot{M}R^{-3}$. If the mass accretion rate is low enough, or the outer radius is large enough that the temperature becomes low enough to allow recombination, the disk becomes unstable and cycles between a state where H is neutral and the mass accretion rate small (quiescence), and a state where H is ionised and the mass accretion rate large (outburst) .

The key aspect of the DIM is its ability to explain the (in)stability of systems, independently of the exact details of transport in the disk. In CVs, novae-likes and dwarf novae divide up according to DIM expectations \cite[an update of this work would be valuable]{smak}. Applied to LMXBs, the DIM predicts all should be transient and fails to account for persistent neutron star LMXBs \cite{vp}.

\subsection{Irradiation}
In LMXBs, the DIM must take into account irradiation heating of the outer disk by high energy photons emitted closer in to the compact object \cite{vp,king}. Irradiation is observationally inferred from optical reprocessing of X-ray bursts, reflection features in X-ray spectra, disk opening angles or higher optical fluxes in persistent LMXBs than in CV disks. The ratio of the irradiation flux to the viscous flux is
$\frac{{\rm F}_{\rm irr}}{{\rm F}_{\rm disk}}\sim \left( \frac{\cal C}{4\pi R^2}\frac{GM \dot{M} }{R_{\rm c}} \right) / \left( \frac{3GM \dot{M} }{8\pi R^3} \right) \sim {\cal C}\frac{R}{R_{\rm c}}$
where the (geometrically diluted) irradiation flux is taken proportional to the potential energy released at the surface of the compact object ($R_c$). With a factor $\sim 1000$ between $R_{\rm c}$ for a white dwarf and a neutron star, reprocessing of only $\sim 0.1$\% of the inner luminosity is enough to make irradiation relevant. Quantitative estimates show this is enough to explain the stability of persistent NS LMXBs \cite{gd}. 

This modification of the DIM to LMXBs raises several questions. First, it is poorly understood how radiation from the inner regions reaches the outer disk: the convex shape of the disk prevents a direct path. Disk warping may help in some cases; Compton diffusion in a large corona is more generic. Second, black hole systems are still farther than NS systems from the stability boundary: does this reflect differences in evolutionary paths or inaccurate estimates of (higher) secular mass transfer rates? 

\subsection{Truncation/Evaporation}
The evolution of an unstable disk can be investigated by numerical techniques, the difficulty is resolving the propagating heat fronts. Studies have shown that reproducing realistic dwarf novae lightcurves requires a substantial weakening (factor 10 or so) of angular momentum transport in quiescence \cite{lasotarev}. Disks with higher values of $\alpha$ ignite too quickly, do not stack up enough matter and lead to frequent, small-amplitude outbursts unlike the observations.

\begin{figure}
  \includegraphics[width=.45\textwidth]{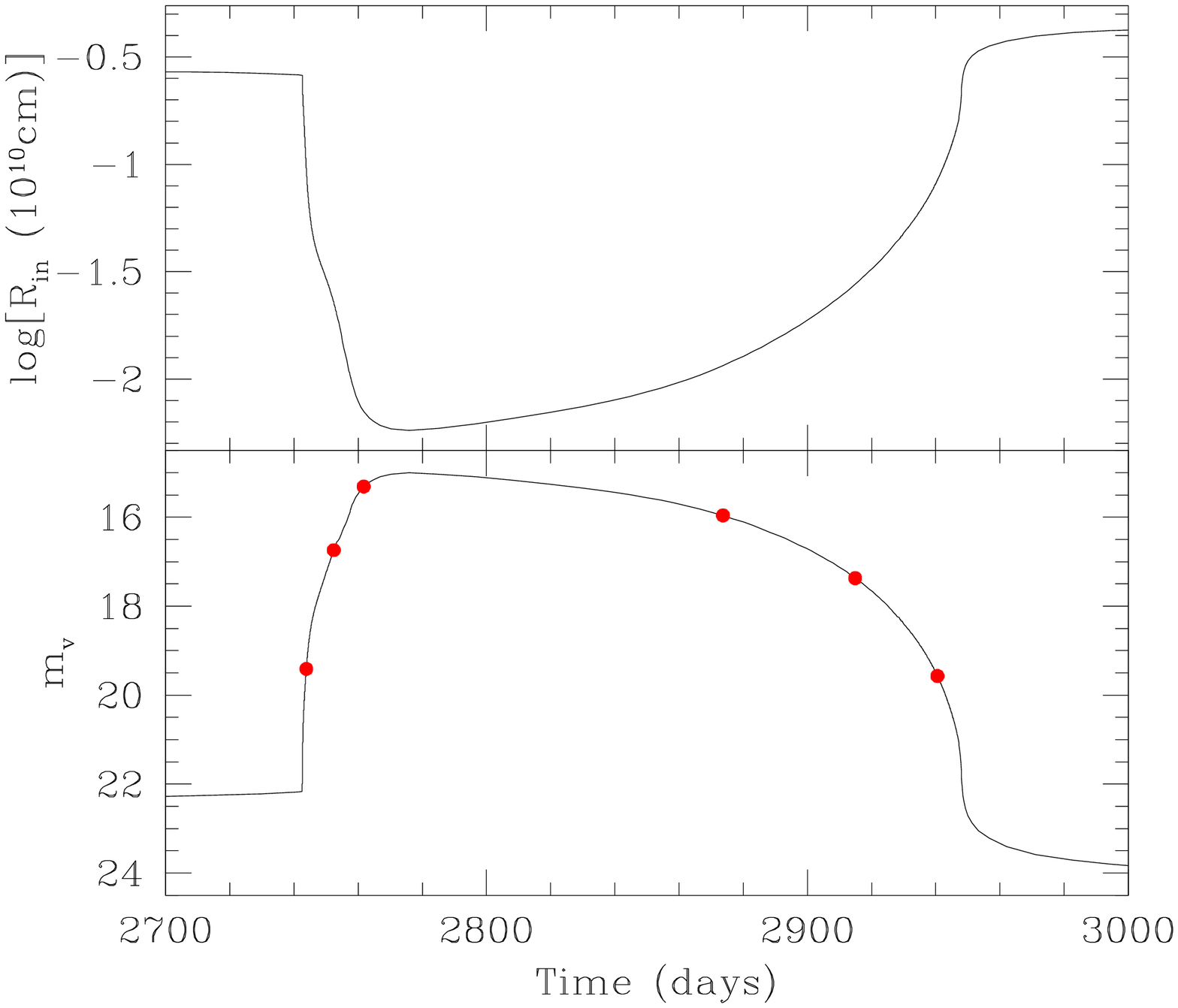}
  \includegraphics[width=.45\textwidth]{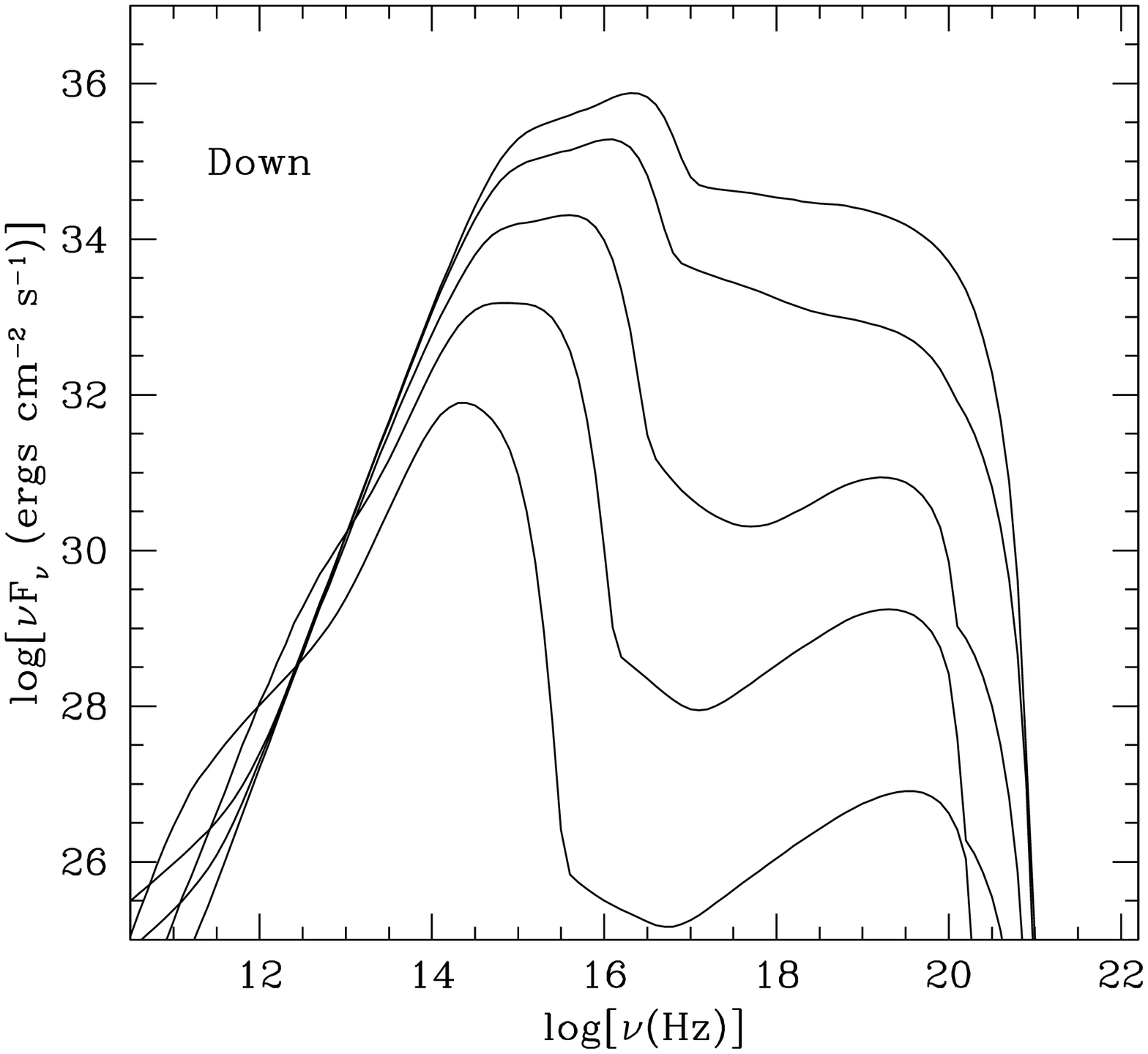}
  \caption{Left: DIM thin disk+ADAF cycle around a 7 $M_\odot$ black hole (top: transition radius, bottom: optical $V$) \cite{dem}. Right: Corresponding X-ray spectrum evolution from near maximum (4 last dots at left) to quiescence (not shown on left) \cite{dem}.}
\end{figure}
In X-ray binaries, the DIM (including irradiation and a lowered $\alpha$ in quiescence) still predicts short cycles inconsistent with observations. Compared to CV disks, LMXB disks extend to much smaller radii where only a small amount of matter suffices to exceed the ignition threshold $\Sigma_{\rm crit}\sim R$. Very low values of $\alpha$ in quiescence mitigate the problem but are not satisfactory: why would such a difference exist at large radii where there is little difference between a CV and a LMXB (WZ Sge is a possible counter-example)? Furthermore, this does not explain how quiescent LMXBs can sustain X-ray luminosities of $10^{31}$~erg~s$^{-1}$ when their inner mass accretion rate must necessarily be lower than the critical one, which is exceedingly low for black holes and neutron stars ($\dot{M}_{\rm cold}\approx 10 (R/R_s)^{2.7} ~\mathrm{kg~s}^{-1}$, independently of $\alpha$).

The thin disk approximation may break down in these regions where new channels open for the energy released by accretion, notably advection \cite{narayanlasota}. For reasons that have yet to be elucidated, the flow takes on a different (necessarily hotter) accretion `solution' (thin disk `truncation' or `evaporation'). With temperatures close to virial, timescales are very short and the flow is stationary. The mass accretion rate dumped on the compact object by the hot flow depends only on the outer thin disk. Much higher mass accretion rates are obtained in quiescence and the resulting outburst lightcurves have the observed long cycles (as a result of trying to fill a leaking bathtub).

Independent support for disk truncation comes from the X-ray spectral evolution of transients, which calls for a two-flow interpretation. The hot flow is a natural candidate for the hard power law seen at low luminosities while the thin disk is the natural for the soft blackbody-ish spectral component that dominates at higher luminosities. As the transition moves in and out, the relative contributions of the two components vary \cite{esin97}.

The truncated, irradiated DIM leads to realistic outburst lightcurves \cite{dubus} (Fig.~1 shows a calculation including an ADAF hot flow+thin disk). Under identical assumptions to CVs for angular momentum transport, the model offers a satisfactory explanation of the stability of systems and of the timescale and amplitudes of outburst cycles. 

\section{Challenges}

\subsection{Upstream: physical processes underlying the DIM}
The DIM calls for reduced angular momentum transport in quiescence. Can this be related to some property of MRI? The MRI needs some fraction of the matter to be ionised in order to develop. If the temperature is too cold then MRI shuts down and accretion stops \cite{menou}. Although even a very small ionisation fraction is sufficient to get the process working, how this can lead to a saturation in quiescence at a tenth the outburst value rather than a simple on/off switch is a puzzle. Alternatively, other transport mechanisms could kick in; in any case some transport must occur as accretion does continue in quiescent CVs as evinced by their X-ray emission.

MRI studies involve heavy numerical simulations that forbid modelling of a thin disk from first principles. Turbulence within the disk height $H$ needs to be resolved so a full disk requires $\sim (R/H)^2$ times the cost of a shearing box simulation. Factoring in the ratio of the viscous to dynamical timescale (another $(R/H)^2$), penalties for radiative transfer and the cost is prohibitive. Global thick disk simulations are easier. Yet, a single zone, vertically resolved simulation including simplified radiative transfer would already provide valuable insights. Where is most of the dissipation occurring ? How does the ionisation change with height ? Can the limit cycle be confirmed ab initio and how does the MRI react to it ? Can better models of disk atmospheres be derived from the simulation, leading to accurate model spectra and line diagnostics in IR/optical/UV (which have more complex troughs and peaks than the deceptively simple double-peaked profile) ? Right now, observations of lines tell us nothing about the conditions in disks.

Another major issue upstream of the DIM is the transition from the thin disk to the hot flow. From the DIM point-of-view, going beyond simple prescriptions would allow detailed predictions of the outburst cycle lengths and luminosities that could be used against, e.g. long term lightcurves of quiescent NS transients. A long term lightcurve of a quiescent {\em black hole} transient in between outbursts has never been obtained (a good target could be 4U1630-47) but would provide interesting insights on how truncation changes as mass accumulates (without truncation the expectation is a steadily increasing X-ray luminosity). 

Models of the transition typically require stitching up in the form of some ad hoc heat term. However, a term in the energy equation representing correlations between velocity and temperature fluctuations has been been hitherto ignored \cite{balbus2}. Using a dimensional $\alpha$-like prescription, can this additional degree-of-freedom be used to investigate what type of smooth transitions between flows are possible ? 

\subsection{Downstream: phenomenology and observations}
The DIM predicts regularly repeating, similar outbursts while observations suggest this to be more the exception than the rule. This must be weighted against the number of cycles on record, incomplete coverage (less of an argument nowadays thanks to the RXTE ASM) or differing bandpasses. Long term monitoring is essential: observations of $>$100 cycles exist for some CVs. Some (many?) transients have rebrightenings and weak outbursts in the years following a major eruption (e.g. XTE J1550-563). Are these the fading echoes of a single event or genuine outbursts as the system moves to a new equilibrium? If truncation occurs up to large radii, the natural state of LMXBs could be stable, low luminosity, cold accreting systems \cite{menounarayan}. Transients would be flukes, temporarily unstable systems due e.g. to fluctuations in the mass transfer rate from the star: no periodicity would be expected.

\begin{figure}
  \includegraphics[width=.32\textwidth]{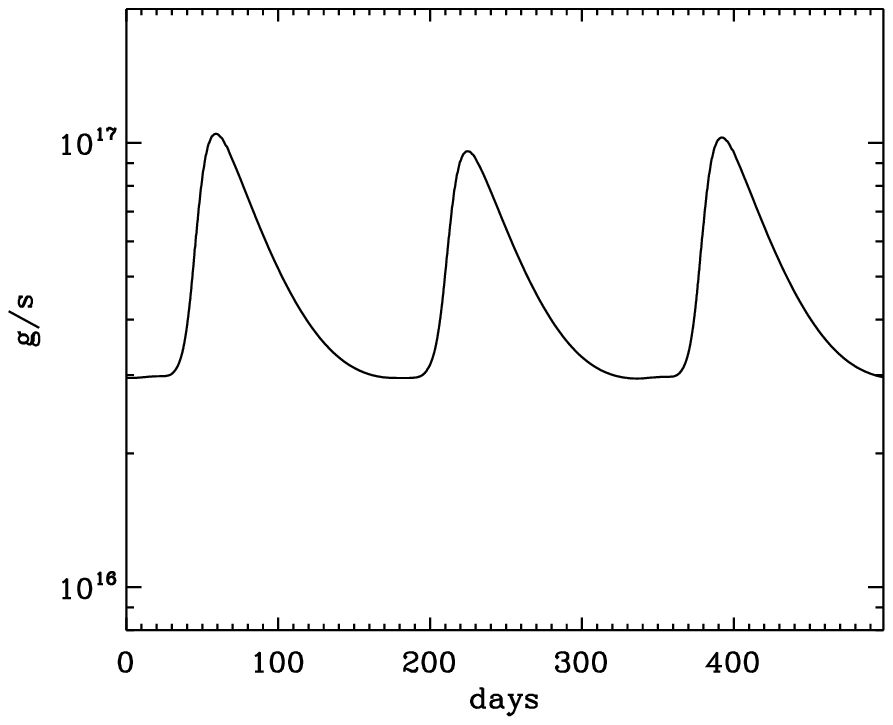}
  \includegraphics[width=.32\textwidth]{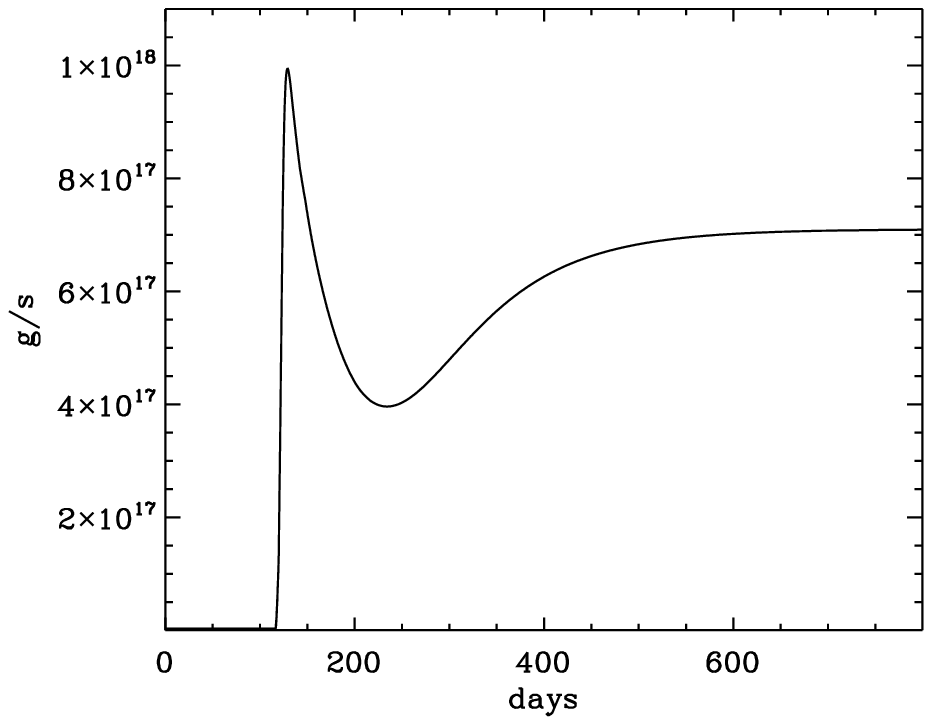}
  \includegraphics[width=.32\textwidth]{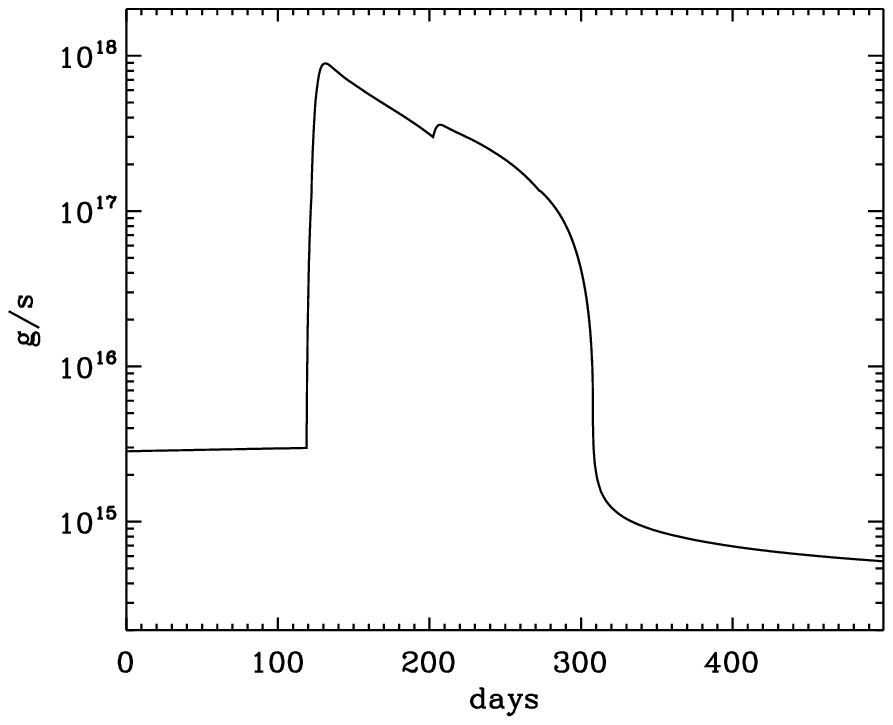}
\caption{Example of possible modifications to the DIM. Left: Long period, small amplitude oscillations appear in an unstable disk when irradiation is kept constant (hence is decoupled from the inner mass accretion rate). Middle: The companion, irradiatied by the initial outburst, increases its mass loss rate after a delay, resulting in a plateau as the disk stabilises. Right: A (radiation-driven?) wind carries off mass from the disk with $R\mathrm{d}\dot{M}_w/\mathrm{d}R\propto \dot{M}_\mathrm{in}$. The glitch occurs when the wind is quenched (change in irradiation spectrum?).}
\end{figure}
Some of the diversity in X-ray lightcurves may reflect differences in inclination hence absorption \cite{narayanmcc}. Still, the DIM does not explain the glitches, plateaus and reflares seen in many outburst lightcurves \cite{ccl}. New understanding may be waiting there (see Fig.~2 for some examples). With small mass ratios, LMXB disks can become elliptic in outburst: are regular reflares \cite[e.g.][]{callanan} related to a shrinking, precessing disk returning to quiescence? Alternatively, do they reflect changes in the irradiation geometry that periodically reignite material? Are plateaus due to enhanced mass transfer from the companion irradiated by the initial outburst \cite{esin2000}? Glitches appear associated with X-ray spectral changes: is the sudden rise in flux a direct consequence of a hardening of X-ray irradiation or because a disk wind is quenched \cite[e.g.][]{cannizzo,mineshige}? Glitches are more prominent in IR and in hard X-rays \cite[e.g.][]{buxton}: are they related to the transition to a hot inflow/outflow (compact jet)?

The DIM is more a model of the outer, thin disk than a model for X-ray emission. Its real test bench is the near-IR to near-UV. Because of the long campaigns needed, less is arguably known on how the lightcurves and spectra vary at these wavelengths than at X-rays. Lightcurves of Aql X-1 or XTE J1550-563 show large amplitude optical outburst that are barely visible (if at all) in the RXTE ASM \cite{maitra,orosz}. Are we missing a lot of the activity in the outer disk region, where the DIM applies? Polarimetry can be used to look for the onset of jet synchrotron emission in IR \cite{dc}. Long term spectroscopy is also needed: changes in the optical spectra with X-ray state seen in GX 339-4 \cite{wu} should be sought in other systems. The far-UV spectrum of XTE J1859+226 changes from soft  to hard (flat in $F_\nu$) during decay \cite{hynes}. The DIM predicts the opposite: in outburst the disk is hot with a distribution  $T\sim R^{-3/4}$ giving a hard UV spectrum; as the outer regions cool the spectrum becomes gradually softer (Fig.~3). Hard X-ray irradiation, not taken into account by the model, penetrates deeper in the disk and could explain the discrepancy. As much attention should be paid in the future to the IR-UV as to the X-ray properties of transients.



\begin{figure}
  \includegraphics[width=.45\textwidth]{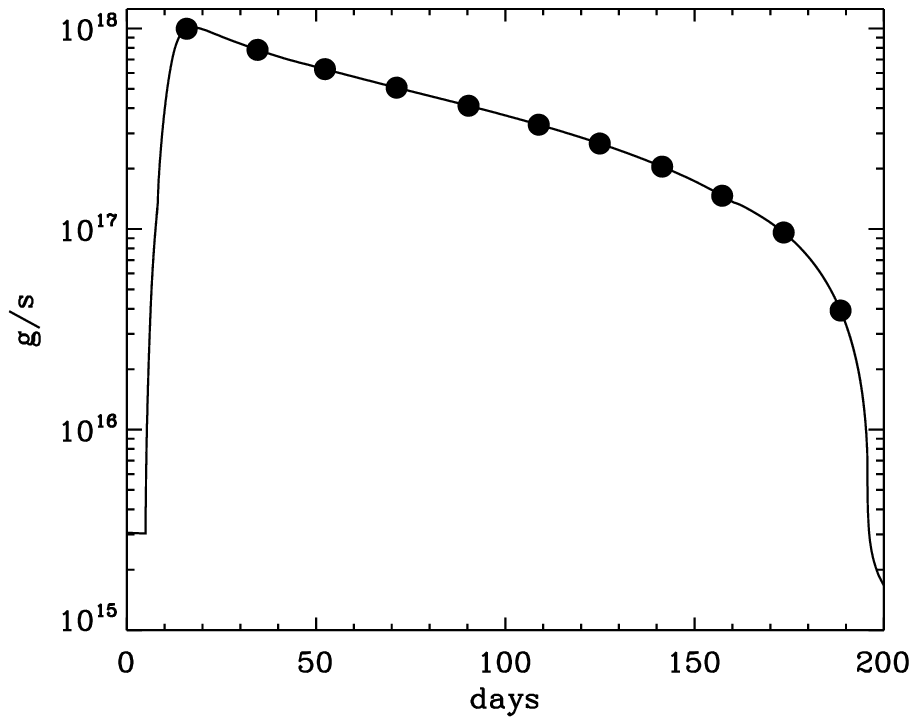}
  \includegraphics[width=.45\textwidth]{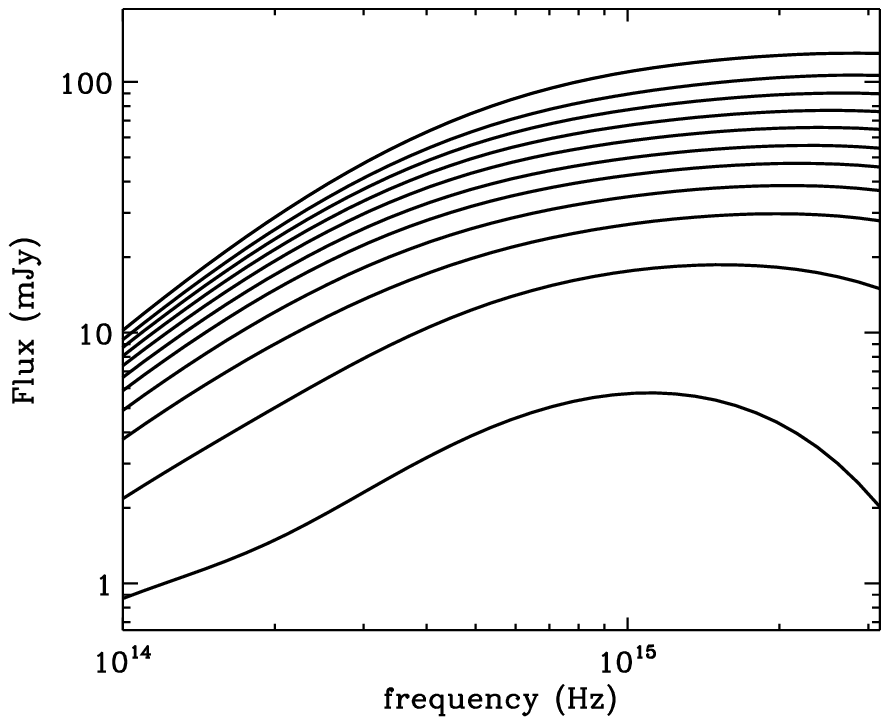}
  \caption{Left: DIM calculation of the mass accretion rate on a 7 $M_\odot$ black hole in outburst \cite{dubus}. Right: Corresponding evolution of the optical/UV SED from peak to quiescence (dots on the left figure identify shown spectra). Compare this to observations of XTE J1859+223 \cite[Fig.3 in][]{hynes}}
\end{figure}




\bibliographystyle{aipproc}   




\end{document}